\documentclass[12pt,a4]{article}
\usepackage{graphicx}
\usepackage{longtable}
\usepackage{amssymb}
\usepackage{amsmath}
\usepackage[all]{xy}

\newcommand{\be}{\begin{equation}}
\newcommand{\ee}{\end{equation}}
\newcommand{\ba}{\begin{eqnarray}}
\newcommand{\ea}{\end{eqnarray}}
\newcommand{\baa}{\begin{eqnarray*}}
\newcommand{\eaa}{\end{eqnarray*}}

\begin{document}

\title{Metamorphosis and Taxonomy of Andreev Bound States}
\author{Cristina Bena$^{1,2}$\\
{\small \it $^1$Laboratoire de Physique des Solides, Universit\'e
Paris-Sud},
\vspace{-.1in}\\{\small \it  B\^at.~510, 91405 Orsay, France}\\
{\small \it $^2$Institut de Physique Th\'eorique, CEA/Saclay,
CNRS, URA 2306},
\vspace{-.1in}\\{\small \it  Orme des Merisiers, F-91191 Gif-sur-Yvette, France}} \maketitle
\maketitle
\begin{abstract}
We analyze the spatial and energy dependence of the local density of states in a SNS junction. We model our system as a one-dimensional tight-binding chain which we solve exactly by numerical diagonalization. We calculate the dependence of the Andreev bound states on position, phase difference, gate voltage, and coupling with the superconducting leads.  Our results confirm the physics predicted by certain analytical approximations, but reveal a much richer set of phenomena beyond the grasp of these approximations, such as the metamorphosis of the discrete states of the normal link (the normal bound states) into Andreev bound states as the leads become superconducting.
\end{abstract}

\section{Introduction}
The Josephson, or superconductor-normal-superconductor (SNS) junctions, reveal some of the most interesting aspects of superconductivity. They are made of  two superconducting (SC) layers and a normal link, and their physics
is dominated by the SC proximity effect by which a gap is induced in the spectrum of the normal link. When taken out of equilibrium, a supercurrent (Josephson current) flows through such a junction \cite{josephson}. The appearance of this
current seems paradoxical, as no current is expected to flow trough a gapped normal link. To solve this, it  has been proposed that discrete
states, called Andreev bound states (ABS), are formed inside the gap \cite{abs}. The supercurrent is carried by
these states, much like a regular
current is carried by the Landauer-Buttiker transmission channels in a ballistic system; thus, the Josephson current is proportional to the number of ABS. This current is also proportional to the phase derivative of the energy
of the ABS states: the larger the sensitivity of an ABS to the phase difference between the superconductors, the larger the Josephson current carried by it.
While these states have been predicted quite a long time ago \cite{abs},
they had only been measured indirectly via transport experiments \cite{absexp1,abs-others}, until their recent direct detection via spectroscopic techniques \cite{abs-saclay,abs-nadia}.

The ABS have been explored theoretically via the Bogoliubov-de-Gennes  (BdG) equations, the Andreev approximation \cite{bdg,affleck}, as well as other approaches \cite{othertheo}. These techniques give the ABS spectrum in various limits,
but cannot capture all the complex effects associated with the finite length
of the link, the eventually asymmetric, imperfect coupling
between the link and the superconductors, the gate voltage, etc. Moreover, the dependence of the density of states in the link on position has been addressed  only in a few works, mostly in relation to the form of the ABS wave-functions \cite{abs-space-dep}.

In this work we calculate the full ABS spectrum, including the limits in which the analytical approximations
described above do not work, by using a lattice one-dimensional tight-binding model \cite{tb} that we diagonalize
numerically.  This procedure allows us to obtain the dependence of the ABS spectrum on position, as well as on
other parameters such as the SC order parameter or the coupling with the leads. By tuning these parameters continuously
we capture the transition between the SC state and the normal state and the way this is manifested in the ABS spectrum.

Furthermore, we elucidate the relation
between the normal state bound states of the link (NBS) and the ABS. The NBS spectrum is made of discrete sharp levels, whose spacing is controlled by the length of the link;
when the normal link is coupled to metallic leads the NBS levels become wider, and the spectrum becomes energy independent for a perfectly-coupled system. Our most important observation is that the ABS are connected to, and evolve from the NBS. Thus we find that the
number of ABS is proportional to the number of NBS falling inside the proximity-induced gap: the larger the ratio between the SC gap
and the NBS level spacing, the larger the number of ABS. Moreover, the spatial structure of the ABS and the NBS is the
same, and there is a one-to-one correspondence between each ABS and NBS in the spectrum!  We confirm this correspondence between the ABS and the NBS spectra by comparing our numerical results with the results of analytical approximations\cite{bdg,affleck}.

The energies of the ABS  are in general different from the energies of
the NBS, and the energy difference depends on the coupling to the leads and the SC gap.
We find that the larger the difference in energy between the ABS and the
NBS, the larger the phase variation of the ABS. When the SC gap is much larger than the coupling with the leads,
the NBS and the ABS have the same energy, and the dependence of the
ABS position on phase is very small.

In section 2 we present the tight-binding Hamiltonian that we use to
describe the system. In section
3 we present the dependence of the density of states (DOS) on position
and energy for non-SC contacts, focusing in particular on the
dependence on the coupling with the leads. In section 4 we analyze the
SC state, focusing on the dependence of the DOS on the coupling with
the leads and the SC gap (section 4.1), on the gate voltage (section
4.2), on the phase difference between the two SC (section 4.3), on
position (section 4.4), and in section 4.5 we describe the DOS in a double-dot configuration. In section 5 we discuss the Andreev approximation (section 5.1),
the ``large-gap'' approximation (section 5.2) and the wide-band
approximation (section 5.3), and compare their outcome to the
numerical results obtained via the tight-binding model.

\section{Model}
\begin{figure}[htbp]
\begin{center}
\includegraphics[width=5in]{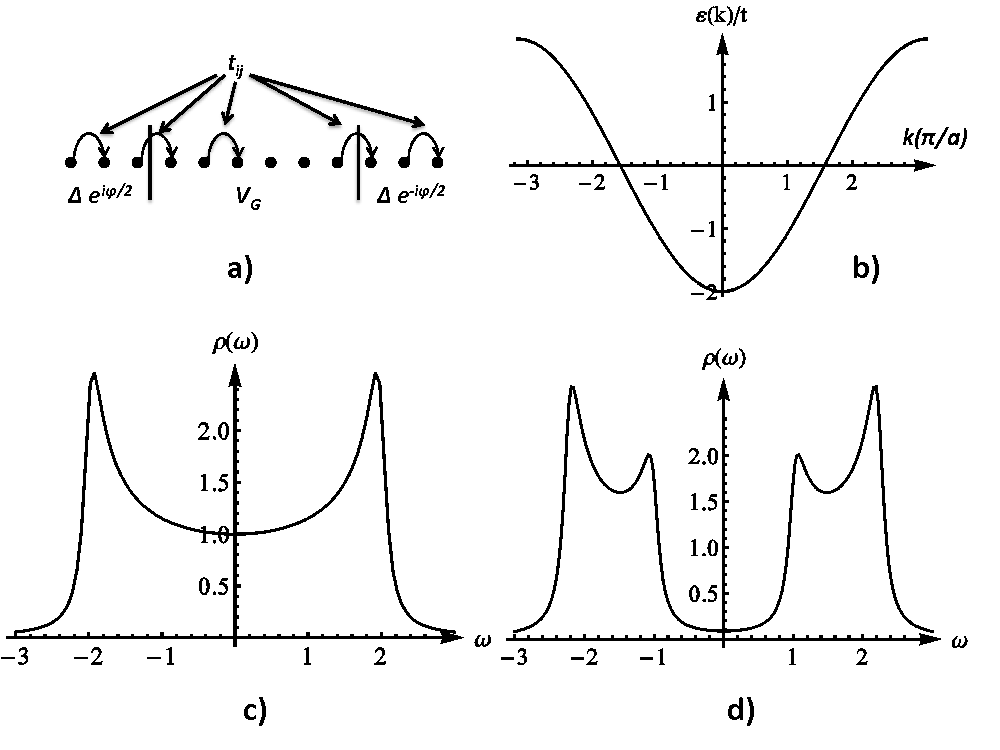}
\caption{\small a) One-dimensional chain. b) The energy dispersion for the infinite uniform system and (c) the corresponding density of states d)The density of states when the gap is non zero ($\Delta=1$). In c) and d) we have set $t=1$, and we have performed the calculations for a uniform system of $100$ sites, with a roughly large $\delta=0.1$}
\label{fig0}
\end{center}
\end{figure}

We model our system as a one-dimensional chain of atoms with inter-atomic distance $a$, and with one electron per atom.  We assume that the number of sites in each lead is $S$, while the number of sites in the link is $N$.  The electrons are allowed to hop between atoms, with a hopping parameters  $t_{i j}$ (see Fig.~\ref{fig0}).  The number of electrons per atom can be modified by en electrostatic potential (gate voltage) which can depend on position, $V_G^i$. The superconductivity in the leads is modeled by an on-site pairing potential $\Delta_i$ \cite{affleck}:
\be
H=\sum_{i=1,2S+N} (\sum_{\alpha=\uparrow,\downarrow}V_{G}^i c_{i\alpha}^\dagger c_{i \alpha}+\Delta_i c_{i \uparrow} c_{i \downarrow})-\sum_{<i,j>} t_{ij} c_i^\dagger c_j+h.c.,
\ee
where $\sum_{< i,j >} $ denotes summing over nearest neighbors.

For simplicity we will assume that for $i \le S$, $\Delta_i=\Delta \exp^{-i \phi/2}$, for $S<i \le S+N$, $\Delta_i=0$, and for  $i>S+N$, $\Delta_i=\Delta \exp^{i \phi/2}$. Thus the superconducting order parameter is uniform in the leads, with a phase difference of $\phi$ between the two leads, and is zero inside the link. An entirely normal system is retrieved when we set $\Delta=0$. For most of the calculations we assume that in the leads the applied gate voltage is zero, while in the link the gate voltage is uniform, $V_G^i=\mu$.

We should note that we do not impose periodic boundary conditions on our chain, so that no confusion arises when incorporating into the model the superconducting gap and the phase difference between the two SCs.

This Hamiltonian can be solved by writing it in a $2(2S+N)\times 2(2S+N)$ matrix form (two states for each sites corresponding to the spin-up/-down) that can be diagonalized exactly numerically using techniques similar to those of \cite{tb}. This diagonalization allows one to obtain the entire eigenvalue spectrum, as well as the corresponding eigenfunctions. Also, by inverting the $(\omega+i \delta)I_{2(2S+N)} - H$ matrix one can obtain  numerically the retarded Green's function matrix, which in turns allows one to obtain the DOS as a function of energy and position.

To compare the validity of this numerical method we compare the numerical results with the corresponding results for an infinite chain.
Thus, for an infinite uniform system with hopping parameter $t$ the Hamiltonian can be written as
\be
H=\int_{-\pi/a}^{\pi/a}[\epsilon(k)-\mu] c_{k}^\dagger c_k
\ee
where $\epsilon(k)=-2 t \cos(k a)$.
The dependence of the DOS on energy for this simple model in the normal state can be calculated:

\be
\rho(E)=-{\rm Im} \int_{-\pi/a}^{\pi/a}G_R(k,\omega)=-{\rm Im}\int_{-\pi/a}^{\pi/a}\frac{1}{\omega-\epsilon(k)+i \delta}\propto\frac{\theta(2t-|\omega|)}{\sqrt{4 t^2-\omega^2}}
\ee

As we can see in Fig.~\ref{fig0} c), this form of the DOS is reproduced by the numerical calculations, even if the size of the considered chain is not too big ($100$ sites). We note that the bandwidth of an uniform infinite system with hopping parameter $t$ is $4t$, with allowed states of energy between $(-2 t, 2t)$. The form of the DOS is well reproduced numerically also in the SC state (se Fig.~\ref{fig0} d).

\section{Normal state}
\subsection{Single impurity}
We start our analysis by reviewing the simple example of an infinite chain with a single localized impurity. We model the impurity as an on-site delta-function potential, $V_0$. Friedel oscillations are generated in the DOS by the scattering of the electrons by the impurity, as described in previous work  \cite{singleimp}. Other types of impurities may be considered, such as  a local modification of the hopping parameter, but the qualitative features presented here do not depend on the exact form of the impurity for energies and positions that are not too small.

\begin{figure}[htbp]
\begin{center}
\includegraphics[width=3in]{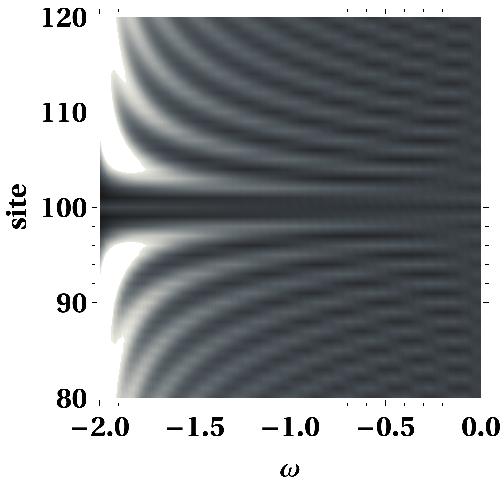}
\caption{\small The density of states as a function of position and energy in a one-dimensional chain with a single impurity. We have taken $t=1$, $V_0=1$, and $\delta=0.04$.}
\label{fig1}
\end{center}
\end{figure}

It is most instructive to plot the LDOS as a function of energy and position, such as in Fig.~\ref{fig1}. The chosen energy range is $(-2t,0)$, below $-2 t$, the bottom of the energy band, the DOS goes to zero. This type of plot allows one to easily identify the Friedel oscillations, and study their dependence on energy. Thus, close to the bottom of the band, the period of the Friedel oscillations is very large, since they arise from scattering of electrons that have similar momenta, as can be seen from Fig.~\ref{fig0}. At energies close to zero, the allowed momentum transfer in a scattering process is close to $2 k_0$, where $k_0=\pi/2a$, yielding Friedel oscillations with a period of $2 a$ (double the atomic lattice constant), and alternating contrast in the DOS on neighboring atoms.

\subsection{Double impurity - quantum dot connected to leads}

We focus next on a one-dimensional chain with two impurities located at the sites $S$, and $S+N$. This can be viewed as a simplified model of a finite-size system  connected to metallic leads via less-than-perfect contacts. One can describe the impurities either by on-site potentials, as we discuss in section \ref{cm}, or by local modifications of the hopping parameter, both yielding similar qualitative results. In what follows we model the impurities as modifications of the hopping parameter. The bad-coupling regime is achieved in this setup when the hopping between the normal link and the leads goes to zero. On the other hand, the good coupling regime is not achieved when the hopping between the link and the leads $t_{imp}$ is infinite, but when it becomes equal to $\sqrt{t_{leads} t_{dot} }$, where $t_{leads}$ and $t_{dot}$ are the bulk hopping parameters in the leads, and in the dot respectively \cite{affleck}. If $t_{dot}=t_{lead}=t$ (the dot and the leads are made of the same material), this prescription yields a perfect coupling for $t_{imp}=t$ (uniform system), as expected.

\begin{figure}[htbp]
\begin{center}
\includegraphics[width=3.5in]{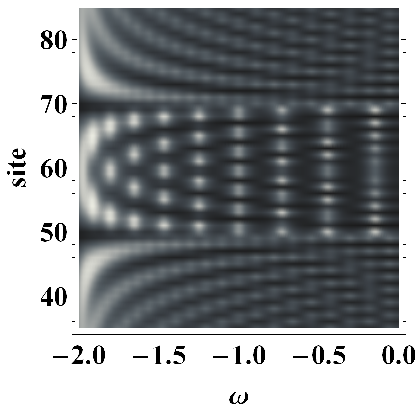}
\caption{\small  The DOS as a function of position and energy in a one-dimensional chain with two impurities. We have taken $S=50$, $N=20$, $\delta=0.05$, $t'=0.2$, and $t=1$.}
\label{fig3}
\end{center}
\end{figure}

We can see in Fig.~\ref{fig3} that the DOS exhibit peaks which correspond to forming of bound states in the finite-size system, and to the quantization of the allowed momenta $k=n \pi/L$. Consequently, the allowed energy values are $E_n=-2 t \cos (n  \pi a/L)=-2 t \cos [\pi  n/ (N+1)]$ , where $N$ is the number of sites. Ignoring the spin, there are $N$ bound states ($n=1,...N$) for the entire bandwidth $(-2t, 2t)$.  Note that if the number of sites is odd, the spectrum contains a zero-energy mode, (obtained when $n=N+1/2$), however, if the number of sites is even, the minimum of the absolute value of the energy is $|E_{N+1/2}|=2 \sin\{\pi/[2(N+1)]\}$. This  dependence of the DOS on position and energy has been studied before in one-dimensional systems, both experimentally and theoretically, for both interacting, and even non-interacting systems \cite{doubleimp}.

\begin{figure}[htbp]
\begin{center}
\includegraphics[width=6in]{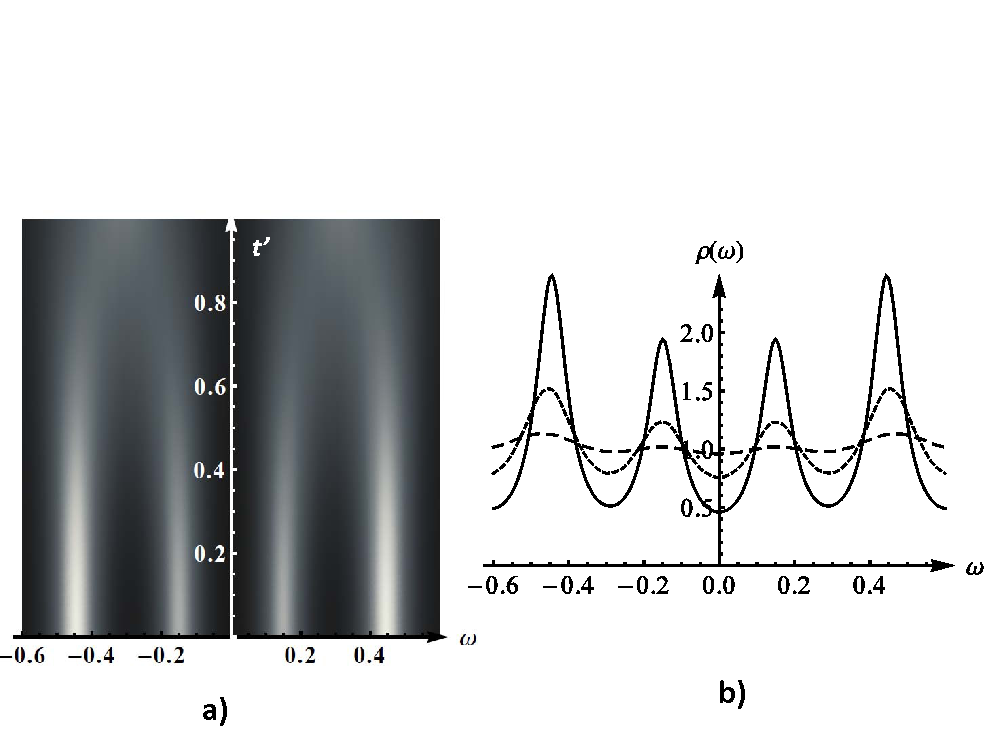}
\caption{\small  a)Two-dimensional plot of the DOS on the site $S+N/2$ as a function of energy and coupling with the leads b) The DOS on one central site as a function of energy for three different values of the coupling with the leads: $t'=0$ (full line), $t'=0.5$ (dashed line), and $t'=0.8$ (large-dashed line). The hopping parameter in the bulk was taken to be $t=1$. The other parameters are $\delta=0.05$, $S=100$, and $N=20$. }
\label{fig4}
\end{center}
\end{figure}

The particular model that we are using allows us to study easily the transition between the Fabry-Perot regime (the central region is coupled almost perfectly to the leads) and the bad-coupling ``Coulomb blockade'' regime \cite{fp}. However, we should note that we cannot capture here the actual Coulomb blockade phenomenon, nor the extra separation between energy levels induced by the Coulomb interactions, since such interactions are not included in our model .
To study the transition between the good-coupling and the bad-coupling regimes we plot the density of state for a site on the link as a function of energy for various values of the coupling with the leads. In Fig.~\ref{fig4} we note that the sharp peaks observed in the bad-coupling regime evolve into wider ones with increasing the coupling, such that the DOS has a quasi-sinusoidal energy dependence in the well-coupled regime.

\subsection{Comparison with the continuous model for an infinite system}
\label{cm}
\begin{figure}[htbp]
\begin{center}
\includegraphics[width=3in]{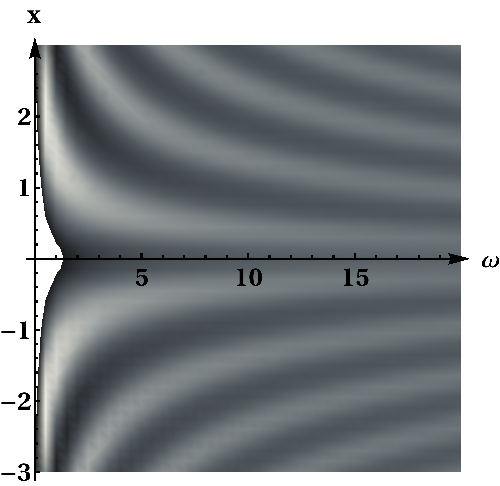}
\caption{\small  The DOS as a function of position and energy in a one-dimensional chain with a single impurity. The impurity potential is set to $V=1$.}
\label{fig6}
\end{center}
\end{figure}

The single- and double-impurity problems can be solved also using the T-matrix approximation  when the length of the wire is infinite. For simplicity we consider a one-dimensional system with a quadratic energy dispersion:
\be
H=\int_{-\pi/a}^{\pi/a}\epsilon(k) c_{k}^\dagger c_k,
\ee
where  $\epsilon(k) \propto k^2$. The quadratic dispersion is a good approximation to the energy dispersion of the tight-binding chain for Fermi energies close to the bottom of the band (almost zero filling). Note that in order to model the tight-binding chain at half filling we need to study a  system with linear dispersion whose spectrum includes right-moving and left-moving particles
\be
H=\int_{-\pi/a}^{\pi/a} \epsilon(k) (c_{R k}^\dagger c_{R k}-c_{L k}^\dagger c_{L k})
\ee
where $c(x)=e^{i k_F x} c_R(x)+e^{-i k_F x} c_L(x)$ and $c_{R k}$ and $c_{L k}$ are the respective Fourier transforms of the fermionic operators.
The retarded Green's functions can be written for this system as a function of energy and position:
\be
G_R(x,\omega)=\int dk \frac{1}{\omega+i \delta-k^2} e^{i k x}
\ee
when we work in units in which the mass of the quasiparticle is equal to $1$. Using this form for the Green's functions, and assuming that the impurity potential is assumed to be a delta-function on-site potential  $V(x)=V\delta(x) c^\dagger(x) c(x)$, the LDOS dependence on position and energy can be determined straightforwardly using the T-matrix approximation \cite{tm1}:
\be
\rho(x,\omega)=-{\rm Im}G_R(x,\omega) G_R(-x, \omega) T(\omega)
\ee
with
\be
T(\omega)=\frac{V}{1-V G_R(0,\omega)}
\ee
In Fig.~\ref{fig6} we plot the resulting LDOS as function of energy and position. Indeed, As expected for a system with quadratric dispersion, at low energy we observe long-wavelength Friedel oscillations, while, at high energy, the DOS exhibits features qualitatively similar to those arising in a tight-binding chain at half filling, such as high frequency, short-wavelength ($k_F^{-1}/2$) Friedel oscillations. We can see that the numerical and analytical result retrieve qualitatively the same form for the Friedel oscillations for energies and positions that are not too small (at low energy the analytical calculation has problems converging, yielding the white cut-out of the LDOS figure).

While not shown here, a similarly good agreement was obtained for the DOS in the presence of two impurities ($V(x)=V\delta(x) c^\dagger(x) c(x)+ V\delta(x-L) c^\dagger(x) c(x)$) using the T-matrix formalism in Ref.~\cite{tm2}.

\subsection{Double quantum dots}

\begin{figure}[htbp]
\begin{center}
\includegraphics[width=3in]{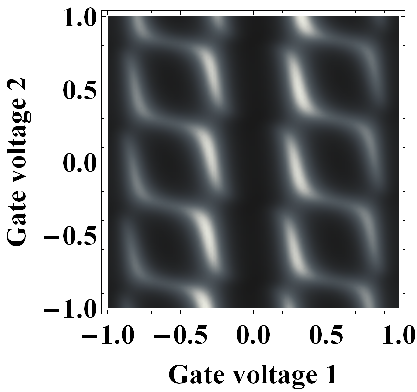}
\caption{\small  The DOS on one dot as a function of the two gate voltages on the two dots. The parameters are $t=1$, $t_{inter-dot}=0.3$, $t'=0.5$.}
\label{fig8}
\end{center}
\end{figure}

We can easily adapt our formalism to study double-quantum-dot systems by modifying the hopping between the sites $S+N/2$ and $S+N/2+1$.
In Fig.~\ref{fig8} we are plotting the DOS on one of the dots as a function of the two gate voltages applied separately on the two dots. If the two dots are completely decoupled, this plot yields a grid of horizontal lines - the states of one dot do not depend on the gate voltage applied on the second dot. However, when the dots are coupled, ``transitions'' occur between the horizontal lines, giving rise to rounded steps. These phenomena have been discussed in detail in the literature \cite{doubledot}, and we confirm here that a tight-binding model recovers a similar form for the dependence of the DOS on the two gate voltages.

\section{Superconducting state}

Our main interest is to investigate the formation of  ABS in the normal link when the leads become superconducting. A vast literature exists  detailing the properties of ABS, their associated wavefunctions, the corresponding Josephson current, etc.\cite{abs,abs-saclay,abs-nadia,bdg,affleck,othertheo}.  While the ABS have been detected experimentally indirectly in  transport measurements \cite{absexp1,abs-others}, a direct spectroscopic measurement has not been done until recently \cite{abs-saclay,abs-nadia}. Most of the ABS theoretical studies have been performed using the Bogoliubov-de-Gennes equations and different approximations \cite{bdg,affleck,othertheo}. Here we use a tight-binding approach, which is more versatile and allows us to modify easily any parameter in the system (coupling with the leads, be it symmetric or asymmetric, phase difference, gate voltage, position), thus giving access to a much richer picture than the one obtained analytically. Moreover, our technique permits to see how the ABS evolve from the bound states existing already in the non-superconducting system, and to study the spatial dependence of the ABS.  Last, but not least, it provides an easy tool to take into account the effects of disorder.

\subsection{The dependence of the ABS on the coupling with the leads, as well as on the superconducting gap}

\begin{figure}[htbp]
\begin{center}
\includegraphics[width=3.5in]{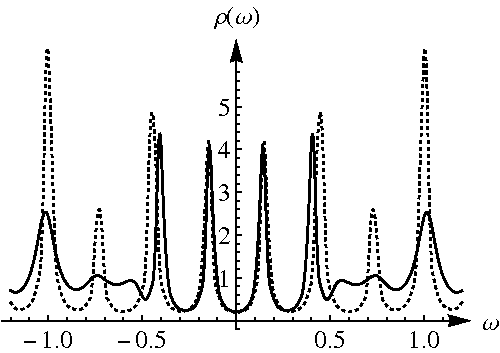}
\caption{\small  The DOS as a function of energy for a system with $S=200$, $N=20$, and $\delta=0.02$. The site for which the DOS is evaluated is $S+N/2=210$. The hopping term $t$ has been set to $1$. The coupling with the leads is set to zero for the dotted line, and to $t'=0.5$ for the full line. The SC gap is set to $\Delta=0.5$.}
\label{fig80}
\end{center}
\end{figure}

\begin{figure}[htbp]
\begin{center}
\includegraphics[width=6.5in,angle=270]{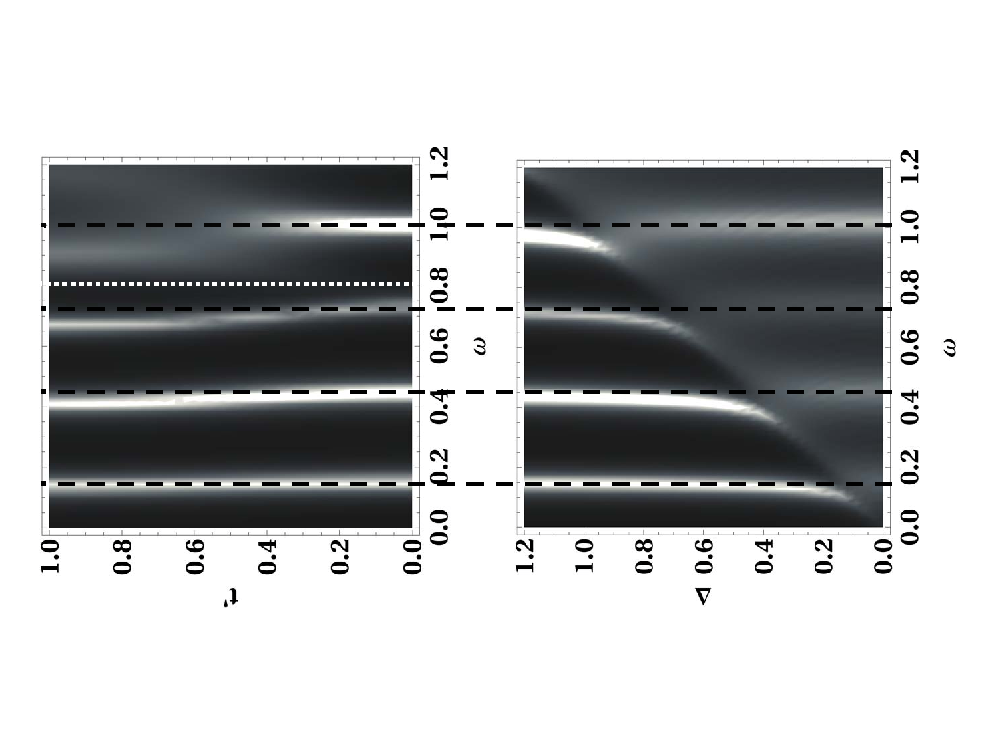}
\caption{\small  The DOS as a function of energy on the x-axis, and, on the y-axis, the coupling to the leads ($t'$) (top panel) and the superconducting gap ($\Delta$) (bottom panel). The coupling with the leads is assumed to be symmetric. We have taken $\Delta=0.8$ for the upper plot, and $t'=0.5$ for the lower plot, $S=150$, $N=20$, and $\delta=0.02$. The site for which the DOS is evaluated is $S+N/2=160$. The hopping term $t$ has been set to $1$. The positions of the NBS are given by the black dashed lines, while the gap is indicated in the upper panel by the white dotted line.}
\label{fig9}
\end{center}
\end{figure}

We begin by considering a one-dimensional chain with two impurities (similar to the one discussed in the previous section)  for which the superconducting order parameter in the leads is non-zero. As described in previous work, a superconducting gap is induced in the spectrum of the normal link, and discrete ABS are formed inside this gap. In Fig.~\ref{fig80} we plot the spectrum of a system with $N=20$ and $S=200$ for an isolated system (zero coupling with the leads) (dotted line), as well as for a coupling of $t'=0.5$ (full line).
When the coupling is non-zero, the spectrum of the system is modified: gap-edge-like features appear at energies close to $\Delta=0.5$, and the position and shape of the NBS peaks change. Thus the peaks inside the gap change position but remain sharp - these peaks correspond to the ABS. On the other hand, the peaks outside the gap are shifted by much smaller amounts, but are smoothed out by the coupling with the leads, same as in the fully normal state. While the energies of the ABS are different from those of the NBS, there seems to be a one-to-one correspondence between each NBS and the ABS.
To understand this issue better, in Fig.~\ref{fig9} we plot the DOS as a function of energy on the x-axis, and, on the y-axis, the coupling with the leads (top panel), and the superconducting gap (bottom panel).
Fig.~\ref{fig9} allows us to understand the evolution of the spectrum between the normal state and the superconducting state. Thus in the bottom panel we note that in the normal state ($\Delta=0$) there are roughly four allowed bound states in the analyzed energy interval, which are broadened by the coupling with the leads. When the value of the gap becomes comparable with the energy of the first state, this state ``enters'' the gap, becoming an ABS. When the gap is increased even more, the second state also enters the gap, etc. For small gaps the ABS are pretty much ``stuck'' to the gap edge, while when the gap increases their energy evolves asymptotically towards a given value. By comparing the lower panel with the upper panel, which describes the DOS as a function of energy and the coupling with the leads, we can see that the positions of the ABS at asymptotically large gap are the same as the positions of the NBS (the NBS spectrum corresponds to $t'=0$ in the upper panel, and the NBS positions are denoted by the black dashed lines).

Fig.~\ref{fig9} - upper panel - describes how the ABS spectrum is modified when the value of the coupling with the leads is modified. The states outside the gap (the position of the gap $\Delta=0.8$ is marked in the upper panel by the white dotted line) become wider when the gap increases, while the states inside the gap have their position slightly modified by the SC coupling, but their widths are unchanged.

From our results one can thus establish a direct connection
between the NBS and the ABS spectra. For example, if there is no NBS in the energy window delimited by the SC gap, no ABS will form, or the ABS will be stuck to the gap edge. Moreover, the number of ABS is directly related to the number of NBS in this energy window. We will discuss in more detail in section \ref{discussion} the analytical interpretation of the correspondence between the NBS and the ABS and in what limits this correspondence can be established.

\subsection{ABS dependence on gate voltage}

\begin{figure}[htbp]
\begin{center}
\includegraphics[width=7in]{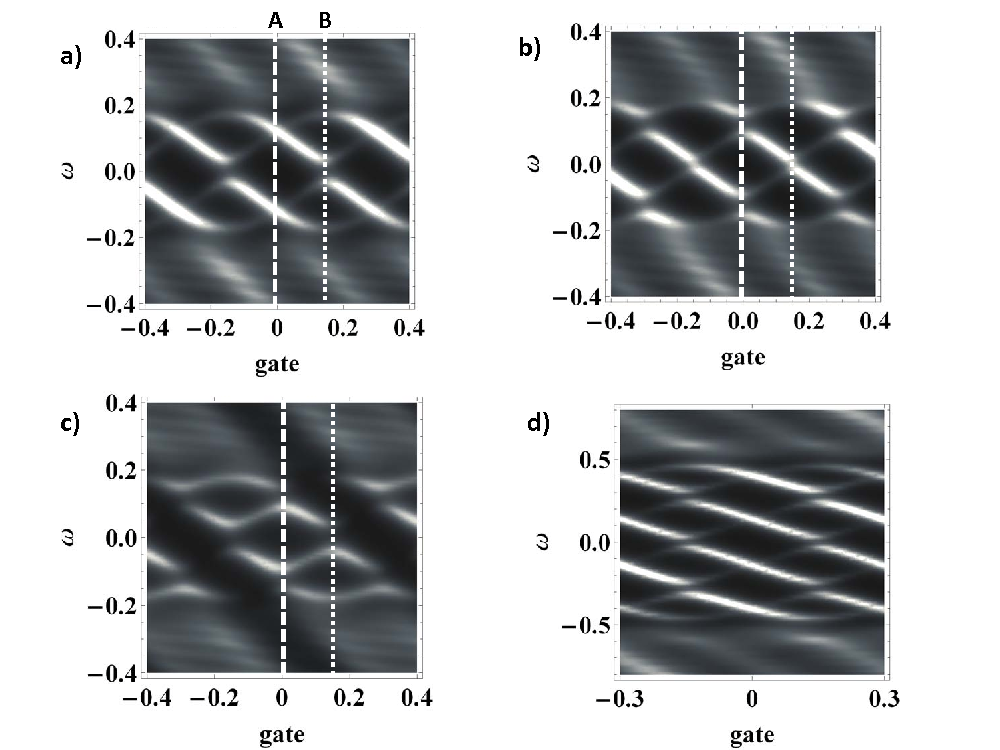}
\caption{\small  The DOS as a function of energy and gate voltage. We have set a)  $N=20$, $S=100$, $t=1$, $t'=0.5$, $\Delta=0.2$, and $\delta=0.02$;  b) same as in a) plus a phase difference between the two SC of $2\pi/3$; c) same as a) plus an asymmetric coupling of the link with the two SCs, $t'_R=0.25$, $t'_L=0.75$; d) same parameters as in a) except for $\Delta=0.5$. The white dashes lines A and B correspond to a gate voltage of $\mu=0$ and respectively $\mu=\pi/N$.}
\label{fig13}
\end{center}
\end{figure}

One of the most interesting aspects of the ABS is their dependence on gate voltage. It is this dependence that allowed one to unambiguously identify the ABS experimentally \cite{abs-saclay,abs-nadia}.
In Fig.~\ref{fig13} a) we plot the DOS as a function of energy and gate voltage for a system of $S=100$ and $N=20$, with $t=1$, $t'=0.5$, and a SC gap of $\Delta=0.2$. For this system the spacing between the normal bound states is $E_L=4 t/N=0.2$.  We note that at $\mu=0$ (along the dashed line in the figure), two ABS form, one at positive energy and the other at negative energy, consistent with the fact that the gap is of the order of $E_L$. Away from $\mu=0$ the states split each into two ABS, which evolve forming bell-shaped/sinusoidal features, and the positive-negative energy symmetry is broken. Outside the gap the position of the states is linear with gate voltage, same as in the normal state; these normal states transform into ABS when they touch the gap edge. Similar features have been described using a simpler model in \cite{abs-saclay}.

We find that, consistent with previous observations, when the value of the gap increases, the number of ABS also increases. Also when the phase difference between the SCs is non-zero, or in the presence of an asymmetric coupling with the two SCs, the ABS at $\mu=0$ split. The ABS also split when the gate voltage is non-zero. We should also note that we do not observe any zero-energy crossing of ABS levels. As described in \cite{abs-saclay}, such crossings, which have been observed experimentally, are the effect of Coulomb interactions and of the lifting of degeneracy between the spin-up and spin-down energy levels. In our tight-binding model such effects are not included and indeed, the crossings do not appear.

\subsection{ABS dependence on phase}

\begin{figure}[htbp]
\begin{center}
\includegraphics[width=7in]{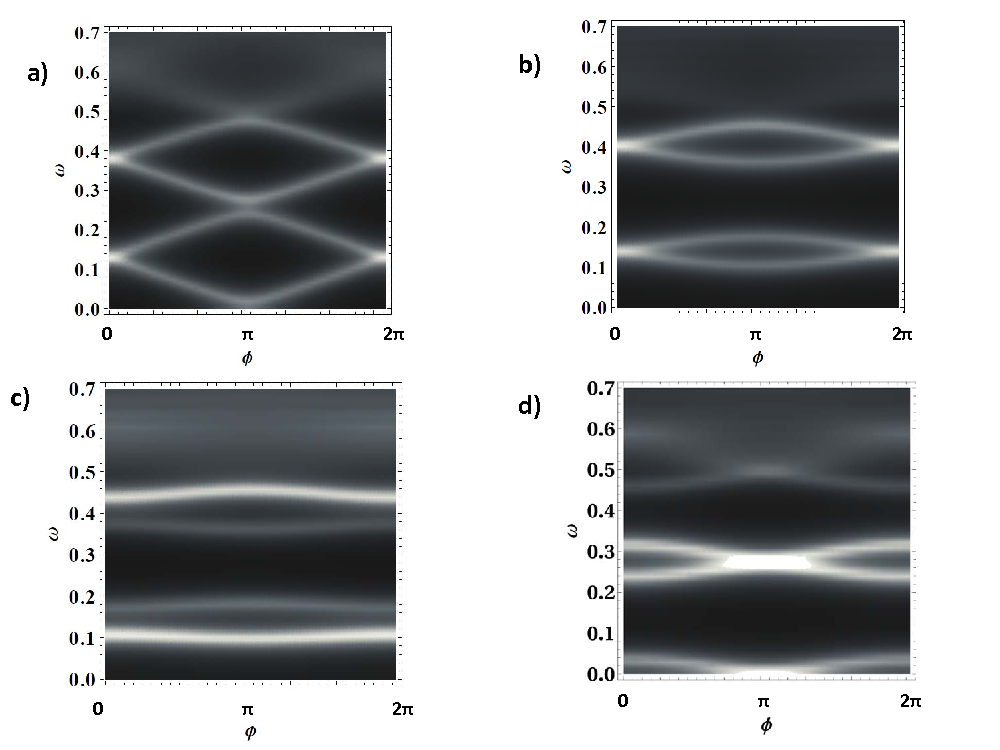}
\caption{\small  The DOS as a function of energy and phase difference on the site $S+N/2$. The parameters are $S=100$, $N=20$, $t=1$, $\delta=0.02$, $\Delta=0.5$, and a)  $t'=1$, b) $t'=0.5$, c) $t'_R=0.25$, $t'_L=0.75$ d)same parameters as a) but for non-zero gate voltage $\mu \approx \pi/(N+1)$. The spectra in a), b) and c) correspond to the line A in Fig.~\ref{fig13}, and the spectrum in d) corresponds to the line B in Fig.~\ref{fig12}.}
\label{fig12}
\end{center}
\end{figure}

We now turn to the study of the dependence of the ABS on the phase difference between the SCs. In Fig.~\ref{fig12} a), b) and c) we plot the DOS as a function of energy and phase difference, for a $\Delta=0.5$ and $\mu=0$, corresponding to the phase dependence of the A spectrum in Fig.~\ref{fig13}. We can see that the two degenerate $\phi=0$ ABS split for non-zero $\phi$. The positions of the ABS vary sinusoidally with $\phi$, the largest amplitude of this variation is achieved when the coupling with the leads is perfect $t'=1$ (a). Moreover, consistent with observations in the previous section, the degeneracy of the two ABS at $\phi=0$ is lifted for an asymmetric coupling with the leads $t'_R=0.25$, $t'_L=0.75$. We should also note that even some of the states outside the gap and close to the gap edge have a non-zero phase dependence, weaker than that of the states inside the gap.

For the generic parameters we have considered, the dependence of the ABS on flux changes qualitatively when the parity of the number of sites in the normal link changes. The difference stems from the effects of this parity on the NBS spectrum. As described in the previous section, if the  number of sites is odd, a NBS zero energy state is formed, while if this number is even, there is no zero-energy NBS.  This is equivalent to saying that the even and the odd spectra are shifted with respect to the zero-energy level by a gate voltage $\mu=\pi/(N+1)$ equal to the difference in energy between the lowest energy eigenstates of the odd and even systems. For a large enough $N$, we can recover an odd spectrum from an even spectrum by a simple gate voltage shift. The spectrum in Fig.~\ref{fig12} (d) evaluated for $N=20$ at $\mu=\pi/N$ (corresponding to the B line in Fig.~\ref{fig13}) should thus be the same as the $\mu=0$ spectrum of a system with $N=21$.


\subsection{ABS dependence on position}

\begin{figure}[htbp]
\begin{center}
\includegraphics[width=6.5in]{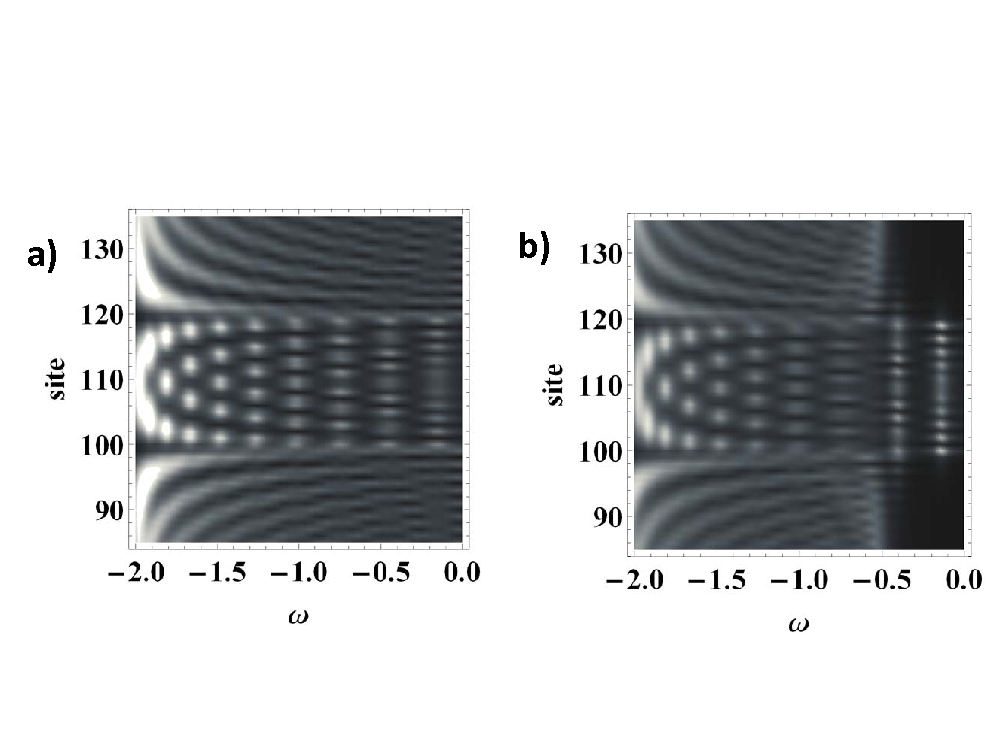}
\caption{\small  The DOS as a function of energy and position for a) $t=1$, $t'=0.5$, $\Delta=0$, $\delta=0.04$, $N=20$, $S=100$, $\phi=0$, $\mu=0$; b)same parameters except for $\Delta=0.5$.}
\label{fig14}
\end{center}
\end{figure}

In Fig.~\ref{fig14} we plot the DOS as a function of energy and position for a) the normal state, and b) the SC state. We observe the SC gap in the leads, the proximity-induced gap in the link and the corresponding ABS. Our analysis reveals  that there is a one-to-one corresponence between the NBS and the ABS, such that for each ABS there is a NBS with the same spatial structure. Thus, each ABS evolves in the presence of the SC leads from an NBS. The main difference between an ABS and its NBS counterpart is the phase dependence of the ABS which is more important for the ABS whose energies differs the most from the NBS energies.

\begin{figure}[htbp]
\begin{center}
\includegraphics[width=7in]{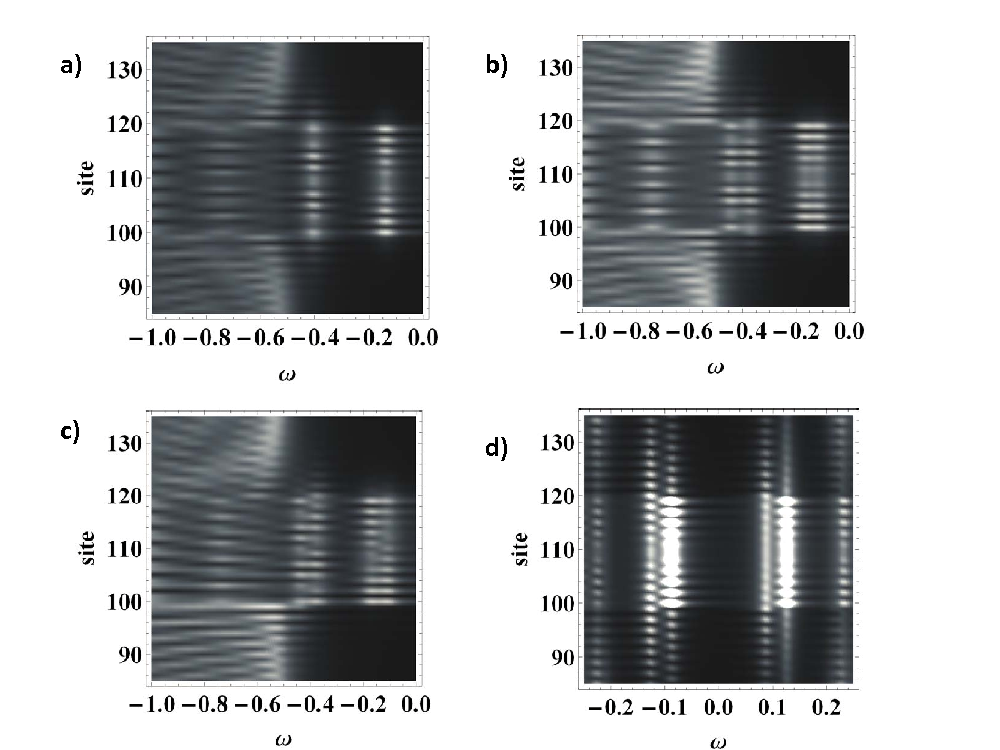}
\caption{\small  The DOS as a function of energy and position. In a) the parameters are identical to the ones in Fig.~\ref{fig14}, in b) we have introduced a non-zero flux $\phi=2 \pi/3$, in c) the coupling with the leads is asymmetric, $t'_R=0.25$, $t'_L=0.75$, while in d) the gate voltage is non-zero $\mu= -0.05$, $\Delta= 0.2$ and $\delta= .01$.
}
\label{fig140}
\end{center}
\end{figure}

In Fig.~\ref{fig140} we zoom in on the lower-energy states: in a) the parameters are identical to the ones in Fig.~\ref{fig14}, in b) we have introduced a non-zero flux $\phi=2 \pi/3$, in c) the coupling with the leads is asymmetric, $t'_R=0.25$, $t'_L=0.75$, while in d) the gate voltage is non-zero $\mu= -0.05$, $\Delta= 0.2$ and $\delta= .01$.
We note that the two states obtained from the splitting of an ABS when $\phi \ne 0$ have  the same spatial structure. However the
two states obtained when the ABS split due to an asymmetric coupling have a different spatial structure.
Furthermore, all the four states (negative and positive energies) arising at $ \mu\ne0$ have the same spatial structure, as it can be seen in Fig.~\ref{fig14}d).
Last, but not least, our analysis confirms that the ABS ``leak'' into the normal leads, and that this leak is larger when the ABS are closer to the gap edge.

\subsection{ABS in double quantum dots}

\begin{figure}[htbp]
\begin{center}
\includegraphics[width=3in]{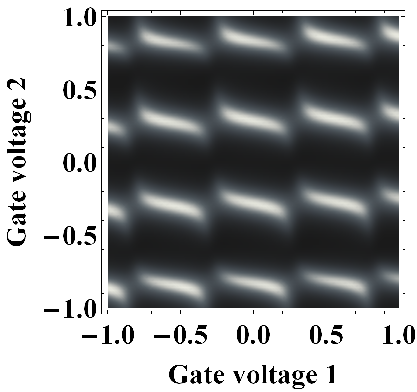}
\caption{\small  The DOS on one dot as a function of the two gate voltages on the two dots. The parameters are set to $t=1$, $t'=0.5$, $t_{inter-dot}=0.5$, $\Delta=0.5$, and $\delta=0.04$.}
\label{fig16}
\end{center}
\end{figure}

By modifying the coupling between the sites $S+N/2$ and $S+N/2+1$ we can split the normal link into two quantum dots.
In Fig.~\ref{fig16} we plot the dependence of the DOS on one dot as a function of the two gate voltages on the two dots. We can see that the picture is very similar to the normal state picture in Fig.~\ref{fig8}.

\section{Discussion}
\label{discussion}

To understand better our results we discuss three particular limits in which the BdG equations can be solved analytically to obtain the energies of the ABS, and we compare these analytical results to our tight-binding results.

\subsection{Low-energy Andreev approximation}

If the lowest-energy ABS form at energies much smaller than the gap, $t/N \ll \Delta \lesssim t$, one is in the regime of validity of the Andreev approximation.  The ABS energies (for perfect Andreev reflection) can be obtained in this limit from the quantization of the phase acquired inside the link \cite{bdg}:

\be
2 \arccos(E/\Delta)-(q_e-q_h)L \pm \phi=2 n \pi
\ee
where $q_e$ and $q_h$ are the corresponding momenta for the electron and the hole with energies $E$ and respectively $-E$, where $E=-2t \cos[(k_F+q_e)a]=2t \cos[(k_F-q_h)a]$.

\begin{figure}[htbp]
\begin{center}
\includegraphics[width=4in]{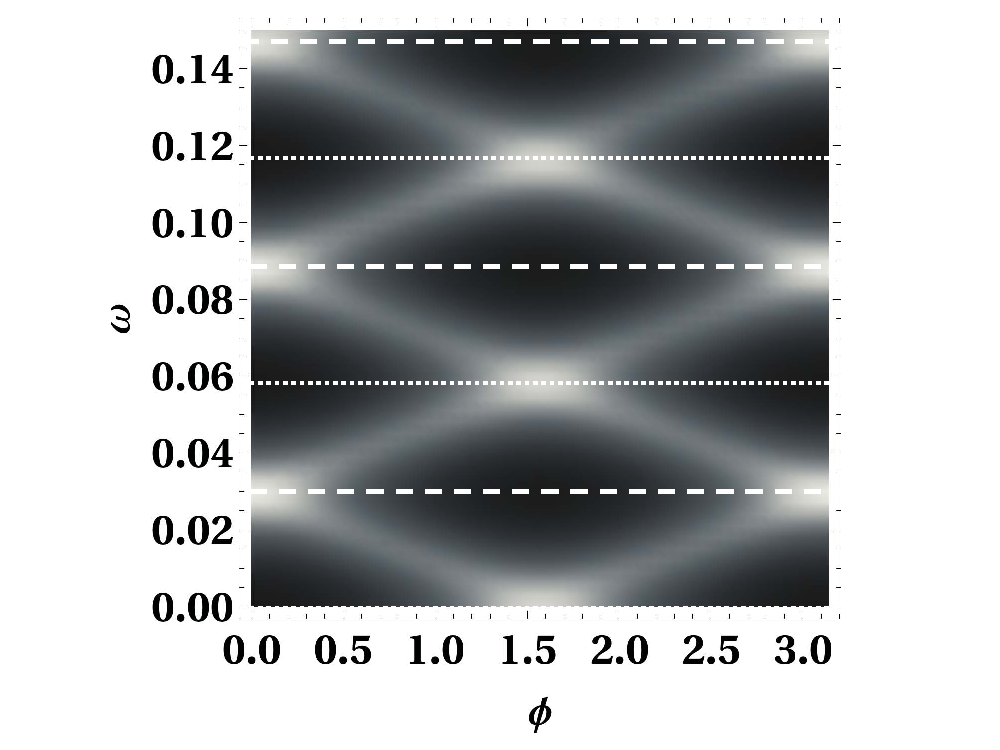}
\caption{\small  Dependence of the DOS on the site $S+1$ on energy and the phase difference between the SCs. The parameters $N=100$, $S=20$, $\Delta=0.3$, $t=t'=1$, and $\delta=0.01$ are chosen so that we are in the Andreev limit. The dashed lines and the dotted lines describe the NBS spectrum for $N=100$ and $N=101$ respectively.}
\label{fig18}
\end{center}
\end{figure}

For energies much smaller than the gap ($E \ll \Delta$, i.e. $\arccos (E/\Delta) \approx \pi/2$), this becomes:
\be
(q_e-q_h) L =\pi(2n-1) \pm \phi.
\ee
Given the band structure presented in Fig.~1, at energies much smaller than $t$, and for chemical potentials $\mu \ll t$, the band can be linearized around the Fermi energy, with $k_F\approx \pi/2a+\mu/(2 t a) $, and $q_e-q_h \approx -  E/(a t)$ independent of $\mu$.
We thus have
\be
E=-t [\pi (2n-1) \pm \phi]/(N+1)
\ee
where we have used the fact that $L=(N+1)a$.
We can compare this quantization relation to the one used to obtain the NBS:
\be
2 k_e L=2 \pi n
\ee
where $E=\mu -2 t \cos(2 k_e a)$, which yields, as discussed in Section 2.1,
\be
\tilde{E}=\mu-2 t \cos {\bigg(} \frac{\pi n}{N+1}{\bigg)}
\ee
The energies of the NBS are different if $N$ is even or odd. Thus, at low energy ($\tilde{E}\ll t$)
\be
\tilde{E}_{odd}=\mu-2 \tilde{n} \pi/(N+1)
\ee
and
\be
\tilde{E}_{even}=\mu-(2 \tilde{n}+1) \pi/(N+1).
\ee
Hence, for $N$ even the NBS spectrum is identical to the ABS spectrum evaluated at $\phi=0$, while for $N$ odd, the spectrum of the NBS is the same as the spectrum of ABS at $\phi=\pi$. This is confirmed numerically; in Fig.~\ref{fig18} we plot the ABS dependence on $\phi$ for $N=100$, $t=t'=1$, and $\Delta=1$. Indicatively, in dashed lines we give the NBS energies for $N=100$ and in dotted lines the NBS energies for $N=101$.

\subsection{Large-gap approximation}

\begin{figure}[htbp]
\begin{center}
\includegraphics[width=4in]{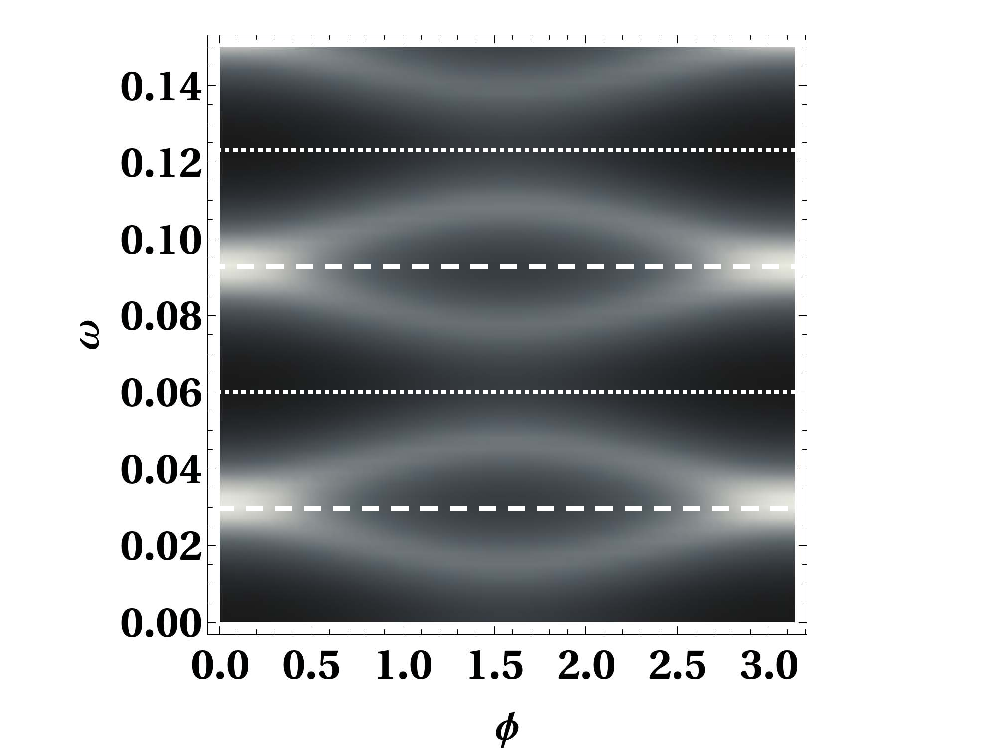}
\caption{\small  Dependence on the DOS on energy and phase for a $N=100$ link, when the system is no longer in the Andreev limit. This is achieved by using the same parameters as the ones in Fig.~\ref{fig18}, and increasing the gap to $\Delta=2$. We can see that the amplitude of the variation with the phase is reduced, and gaps open at $\phi=\pi$. Same as above, the dashed lines and the dotted lines describe the NBS spectrum for $N=100$ and $N=101$ respectively.}
\label{fig17}
\end{center}
\end{figure}

In what follows we analyze the limit in which the gap is much larger than the hopping parameter $t$. This limit is quite unphysical, but it is interesting to study it analytically, since it reveals the mechanism via which the ABS can be identified with the NBS for a specific range of parameters. We begin this analysis  by a numerical study, focusing on the lowest-energy ABS, whose energy is much smaller than $t$. When $t$ and $\Delta$ are comparable, the system is described quite well by the formalism presented in the previous section. A typical dependence on phase in this limit has been presented in Fig.~\ref{fig18}. When $\Delta$ is increased above $t$, we can see in Fig.~\ref{fig17}, that, for $N=100$, the amplitude of the variation of the ABS with the phase is decreased, and a gap opens at $\phi=\pi$. When the gap becomes even larger, the amplitude of the oscillations goes to zero, the ABS become phase-independent and their energies converge towards the NBS energies. While not depicted here, for $N$ odd, the ABS energies also converge at large gap towards the energies of the NBS, which, when $N$ is odd, are identical to the small-gap ABS energies at $\phi=\pi$.

We can understand the origin of this phenomenon by
finding the ABS energies exactly for an extra-small system with one SC site in each lead, and one normal site. When an expansion of the energies of the ABS is performed at large gap to first order in $t/\Delta$,  we find the allowed energies to be independent of the phase difference, and their values to be given by $E= \pm \mu,\pm \Delta$. Indeed, the $E=\mu$ is the energy of the NBS for a single site at chemical potential $\mu$. The extra energy in the spectrum, $E=-\mu$, has a negligible weight, so it does not give rise to a peak in the DOS.

\subsection{Wide-band approximation}

We can also work in the limit of  $\omega \ll \Delta \ll t$. In this limit one can use the wide-band approximation \cite{alfredo,chinese} which allows one to take into account the superconducting leads by a adding to the Green's function matrix of the normal link a self-energy resulting from integrating out the leads. In the limit $\omega \ll \Delta$, the formalism presented in \cite{alfredo,chinese} yields  self energy contributions for the right and left leads in the form of $2 \times 2$  matrices in the BdG basis, whose only non-zero elements are off-diagonal and are given by $\pm  \Gamma_{R/L} e^{i \phi_{R/L}}$. The DOS can be obtained exactly in this limit for a link made of one site or two sites. For a one-site link we find:
\be
E=\pm \Gamma \cos(\phi)
\ee
while for a two-site problem
\be
E=\pm\sqrt{\Gamma^2 + t^2 \pm 2 \Gamma t \sin(\phi/2)}
\ee
We can see that in this limit the energy of the ABS is dominated by the coupling with the leads, rather than by the intrinsic energies of the NBS.

\section{Conclusion}

We have analyzed the spectrum of an SNS junction, focusing in
particular on the dependence of the DOS on energy and position. We have used a tight-binding numerical method that
has allowed us to study a broad range of parameters, and thus to go beyond previous
analytical approaches. Our calculations yield similar features to
those recently measured in SNS junctions with carbon nanotubes \cite{abs-saclay} and graphene quantum dots \cite{abs-nadia}, which further confirms that the generic parameters used in our numerical calculations are realistic.  Since these parameter values do not correspond to any limit accessible via analytical approximations, we believe that the full physics of the ABS in such a system can only be obtained numerically.

Furthermore, our analysis reveals a one-to-one
correspondence between the NBS states and the ABS states, whose spectra become identical when the
the SC gap is much larger than the bandwidth of the normal link. In such a limit one
may hope to obtain access to the NBS spectrum (which is in practice masked by the coupling with the
leads) via a measure of the ABS spectrum.

While not presented here, we have also verified that our numerical
tight-binding results are consistent with results obtained by solving the corresponding BdG equations numerically. The analysis required to obtain such solutions is however much heavier, and less elegant, and we believe that the tight-binding method presented here is faster and more versatile.  It would be interesting to generalize our calculation to include the effects of disorder, and to test the stability of the ABS for a disordered system.

Last, but not least, our analysis provides a criterion for the
spectroscopic observation of the ABS in a given clean ballistic
system, based on the energy spacing of the NBS. Thus, if the distance between the NBS levels is much smaller than the SC gap, the ABS may be too close to be resolved in energy. Reversely, if the distance between the NBS is very large, the ABS will
tend to stick to the gap edge, except when the NBS approach the Fermi level and one pair of ABS at a time will be resolved inside the SC gap.

{\bf Aknowledgements}
\\
We would like to thank G. Montambaux, P. Joyez, J.-D. Pillet, and M. Goffmann for interesting discussions. This work has been funded by the ERC Starting Independent Researcher Grant 256965 NANO-GRAPHENE.


\begin{thebibliography}{99}

\bibitem{josephson} B. D. Josephson, Rev. Mod. Phys. {\bf 46} 251 (1974); P.-G. de Gennes, Superconductivity of Metals and Alloys, Benjamin, New York, 1966; A. Griffin and J. Demers, Phys. Rev. B {\bf 4}, 2202 (1971); E. Blonder, M. Tinkham, and T. M. Klapwijk, Phys. Rev. B {\bf 25}, 4515 (1982).

\bibitem{abs} I. Kulik, Soviet Physics JETP-USSR {\bf 30} 944 (1970); C. W. J. Beenakker, cond-mat/0406127 (2004), Transport Phenomena in Mesoscopic Systems, edited by H. Fukuyama and T. Ando (Springer, Berlin, 1992); A. F. Andreev, Sov. Phys. JETP {\bf 19}, 1228 (1964); C. W. J. Beenakker, Phys. Rev. Lett. {\bf 67}, 3836 (1991).

\bibitem{absexp1} M. L. Della Rocca, et al., Phys. Rev. Lett. {\bf 99}, 127005 (2007).

\bibitem{abs-others} R. S. Deacon,  et al., Phys. Rev. Lett. {\bf 104}, 076805 (2010).

\bibitem{abs-saclay}J.-D. Pillet et al., Nat. Phys. {\bf 11}, (2010).

\bibitem{abs-nadia} T. Dirks, et al., Nat. Phys. {\bf 7}, 386 (2011)

\bibitem{bdg} A. Furusaki and M. Tsukada, Phys. Rev. B {\bf 43} 10164 (1991); M. Hurd and G. Wendin, Phys. Rev. B, {\bf 49}, 15258 (194); A. Richter, P. Baars, U. Merkt, Physica E {\bf 12}, 911 (2002).

\bibitem{affleck} I. Affleck, J.S. Caux and A. M. Zagoskin, Phys. Rev B {\bf 62}, 1433 (2000).

\bibitem{othertheo} T. Meng, S. Florens, and P. Simon, Phys. Rev. B {\bf 79}, 224521 (2009); J. Caux, H. Saleur, and F. Siano, Phys. Rev. Lett. {\bf 88}, 106402 (2002); R. Fazio, F. Hekking, and A. Odintsov, Phys. Rev. Lett. {\bf 74}, 1843 (1995);  	E. Perfetto, G. Stefanucci, and M. Cini, Phys. Rev. B {\bf 80}, 205408 (2009); J. Bauer, A. Oguri and A. C. Hewson, J. Phys. Condens. Matter {\bf 19} 486211 (2007); 11A. Martin-Rodero, F. J. Garcia-Vidal, and A. Levy Yeyati, Phys. Rev. B {\bf 72}, 554 (1994); A. Levy Yeyati, A. Martin-Rodero, and F. J. Garcia-Vidal, ibid. {\bf 51}, 3743 (1995); J. C. Cuevas, A. Martin-Rodero, and A. Levy Yeyati, ibid. {\bf 54}, 7366 (1996).

\bibitem{abs-space-dep} A. Nakayama, J. Appl. Phys. {\bf 91}, 7119 (2002).

\bibitem{tb} A. Siber, Am. J. Phys. 74, 692 (2006).

\bibitem{singleimp} I. Schneider and S. Eggert  Phys. Rev. Lett. {\bf 104}, 036402 (2010); P. Kakashvili, H. Johannesson, and S. Eggert, Phys. Rev. B {\bf 74}, 085114 (2006).

\bibitem{doubleimp} G. Buchs, D. Bercioux, P. Ruffieux, P. Groning, H. Grabert, and O. Groning, Phys. Rev. Lett. {\bf 102}, 245505 (2009); I. Schneider, A. Struck, M. Bortz and S. Eggert, Phys. Rev. Lett. {\bf 101}  206401 (2008); S. G. Lemay et. al., Nature {\bf 412}, 617 (2001).

\bibitem{fp} See e.g. K. Grove-Rasmussen, H.I. Jorgensen, P.E. Lindelof, Physica E {\bf 40} 92 (2008).

\bibitem{tm1} J. M. Byers, M. E. Flatte, and D. J. Scalapino, Phys. Rev. Lett. {\bf 71}, 3363 (1993); M. I. Salkola, A. V. Balatsky, and D. J. Scalapino, Phys. Rev. Lett. {\bf 77}, 1841 (1996); W. Ziegler et. al., Phys. Rev. B {\bf 53}, 8704 (1996).

\bibitem{tm2}  X.-L. Song, Z.-Y. Zhao, Y. Wang Y.-M. Shi, Journal of Shanghai University ( English Edition ), {\bf 7(4)}, 361 (2003).

\bibitem{doubledot} W. G. van der Wiel et. al., Rev. Mod. Phys. {\bf 75} 1 (2003).

\bibitem{alfredo} E. Vecino, A. Martin-Rodero, and A. Levy Yeyati, Phys. Rev. B {\bf 68}, 035105 (2003).

\bibitem{chinese} Y. Zhu, Q.-f Sun, and T.-h Lin, J. Phys. Condens. Matter {\bf 13} 8783 (2001).


\end{thebibliography}
\end{document}